# Challenges in Cepheid Evolution and Pulsation Modeling

*Joyce A. Guzik*
*Los Alamos National Laboratory*
*XTD-NTA, MS T-082, Los Alamos, NM 87545*
*joy@lanl.gov*

*Jason Jackiewicz*
*Department of Physics and Astronomy*
*New Mexico State University*
*Las Cruces, NM 88003*
*jasonj@nmsu.edu*

*Nancy R. Evans*
*Harvard Smithsonian Center for Astrophysics*
*Cambridge, MA 02138*
*nevans@cfa.harvard.edu*

**Abstract**

Cepheids have long been used as standard candles to determine distances around the Milky Way and to nearby galaxies. A discrepancy still remains for Hubble Constant determinations using Cepheids vs. the cosmic microwave background or calibrations to the tip of the red-giant branch. Therefore, refinement of Cepheid period-luminosity relations continues to be an active topic of research.

Beyond their utility for the cosmic distance scale, Cepheids are also important 'laboratories' for stellar physics. This paper explores outstanding questions and current areas of research in Cepheid evolution and pulsation modeling. We examine the discrepancy between Cepheid masses determined from pulsation properties and binary orbital dynamics and those determined using stellar evolution models. We review attempts to resolve the discrepancy by including rotation, convective overshooting, and mass loss. We also review the impact of uncertainties in the $^{12}C(\alpha,\gamma)^{16}O$ and triple-$\alpha$ nuclear reaction rates on Cepheid evolution and the extent of 'blue loops' in the Hertzsprung-Russell diagram during which Cepheids are undergoing core helium burning. We consider implications for Cepheids of stellar opacity revisions suggested in light of findings for the Sun and other types of variable stars.

We highlight the opportunity to use the 1-D open-source MESA stellar evolution code and the MESA radial stellar pulsation (RSP) nonlinear hydrodynamics code to investigate changes in input physics for Cepheid models. We touch on progress in 2-D and 3-D stellar modeling applied to Cepheids. Additional areas in which Cepheid models are being tested against observations include: predicting the edges of the Cepheid pulsation instability strip; predictions for period-change rates and implications for instability strip crossings; explaining period and amplitude modulations similar to the Blazhko effect of RR Lyr stars and the origins of additional periodicities that may be non-radial pulsation modes; discovering what can be learned from Cepheid observations in X-ray, ultraviolet, and radio wavelengths. We also show a few examples of Cepheid light curves from NASA TESS photometry.

## 1. Introduction

Classical Cepheids are pulsating variable stars. The prototype delta Cep was discovered by J. Goodricke in 1784. These stars are 4 to 15 times the mass of the Sun and are fusing helium into carbon and oxygen in their cores. They have spectral types F through K and temperatures at the photosphere of around 6000 K. Cepheids lie on an 'instability strip' in the Hertzsprung-Russell (H-R) diagram that intersects with the RR Lyrae instability region on the horizontal branch and delta Scuti instability region on the main sequence (see, e.g., Bhardwaj 2020).

Figure 1 shows an evolution track on the H-R diagram for a 5 solar-mass ($M_{sun}$) model calculated using the MESA code (see section 3). The model starts from the zero-age main sequence where core hydrogen burning begins and is evolved to the point (red circle) where the model is ascending the asymptotic giant branch (AGB) following core helium-burning. The model crosses the Cepheid pulsation instability strip for the first time as it evolves toward lower effective temperatures (redward) after core hydrogen exhaustion on its way to ascending the red giant branch (RGB). At the onset of core helium burning, the model makes an excursion to the left ('blue loop'). Following core helium exhaustion, the model then ascends the AGB, where substantial mass loss occurs. Such a star will eventually end its life as white dwarf. During the 'blue loop' excursion, the model barely reaches the Cepheid pulsation instability region, but can remain there for a while since core helium burning is a relatively long-lived evolutionary phase.

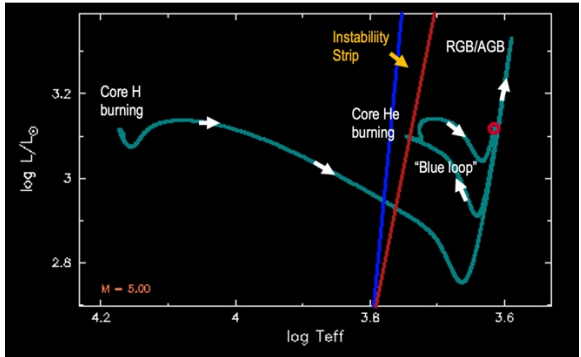

Figure 1. Hertzsprung-Russell diagram showing evolution track of 5 $M_{sun}$ model calculated using MESA. This model crosses the Cepheid pulsation instability strip on the way to becoming a red giant and (barely) reaches the Cepheid pulsation instability strip during its 'blue-loop' excursion while core helium burning.

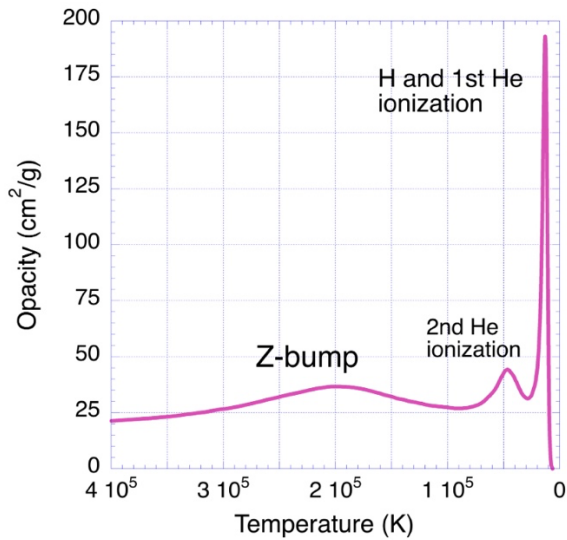

Figure 2. Opacity vs. temperature in the envelope of a stellar model. Cepheid, RR Lyrae, and delta Scuti star pulsations are driven by the 'kappa effect' operating in the 2nd helium ionization region at around 50,000 K.

Cepheid pulsations are driven by the so-called 'kappa' or opacity valving mechanism in the stellar envelope at temperatures of around 50,000 K where helium is becoming fully ionized. Cepheids can pulsate in the radial fundamental mode, first overtone mode, or, more rarely, the second overtone mode, and sometimes show more than one pulsation mode simultaneously. Some Cepheids also are suspected to pulsate additionally in one or more non-radial modes.

The pulsation periods of Cepheids range from 3 to around 100 days. Figure 3 shows the number of Cepheids vs. period from the catalog of 509 Galactic Cepheids at the David Dunlop Observatory website: https://www.astro.utoronto.ca/DDO/research/cepheids/cepheids.html.

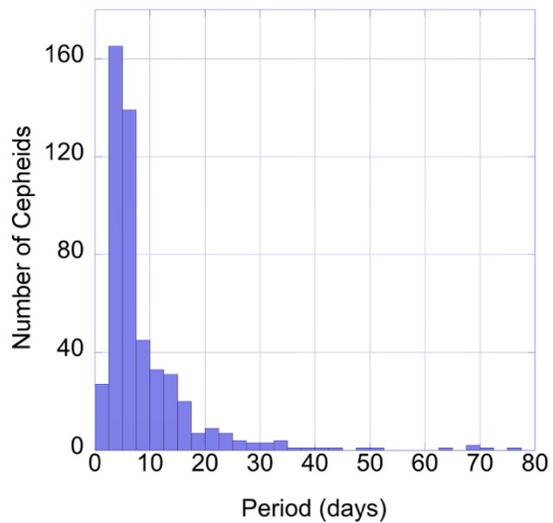

Figure 3. Number of stars vs. period for 509 Galactic Cepheids from database hosted by David Dunlop Observatory, Ontario, Canada.

Cepheids are important because they show a period-luminosity (P-L) relation used to measure distances to nearby galaxies, calibrate distances for brighter 'standard candle' objects such as Type I supernovae, derive the Hubble constant, and set the distance scale of the universe. The P-L relation was first discovered by Henrietta Leavitt in 1908 using observations of Cepheids in the Small Magellanic Cloud. In the 1920s, Edwin Hubble discovered Cepheids in the Andromeda galaxy, proving that this galaxy lies far outside the Milky Way. The P-L relation itself must be calibrated by determining distances to nearby Milky Way Cepheids independently.

Figure 4 shows absolute magnitudes vs. period for 509 Galactic Cepheids from the DDO catalog, illustrating the P-L relation. Both the fundamental-mode pulsator sequence, and the brighter 1st-overtone sequence are evident.

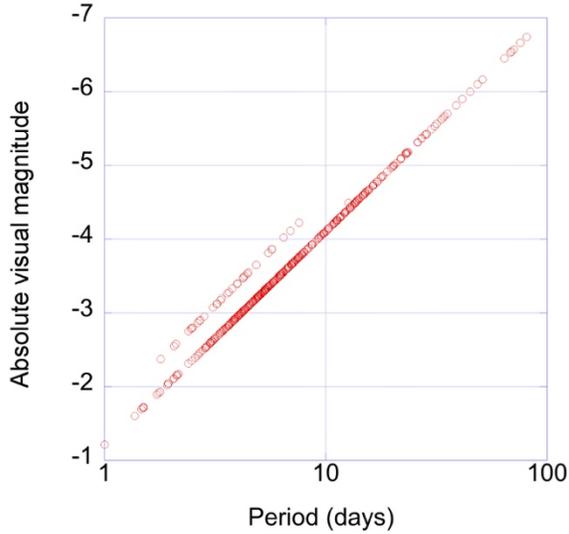

Figure 4. Absolute visual magnitude vs. period for 509 Cepheids from David Dunlop Observatory catalog.

While the Hubble constant has now been determined to be around 70 km/sec/Mpc, determinations of the Hubble constant using Cepheids and those calibrated using the tip of the red giant branch or the cosmic microwave background (CMB) range between 67 and 74 km/sec/Mpc. The uncertainties for each method do not overlap, and there is about a four-sigma discrepancy between the Cepheid and CMB determinations (Freedman 2021). Parallaxes from the Gaia mission (Vallenari et al., Gaia Collaboration 2022) should help to resolve these discrepancies.

Cepheids are also important as 'laboratories' to test stellar physics and modeling under extreme conditions not attainable in laboratories on Earth. Some of the physics areas being explored using Cepheid models include:
- Opacities
- Equation of State
- Nuclear reaction rates
- Convection
- Mass loss
- Rotation
- Nonlinear hydrodynamics
- 2-D and 3-D simulations

In this paper we review some of the challenges and outstanding questions in modeling classical Cepheid evolution and pulsation. Some reviews that also cover observational aspects can be found in Buchler (2009), Daszyńska-Daszkiewicz (2009), Marconi (2017), Bhardwaj (2020), and Plachy et al. (2021).

## 2. Cepheid Mass Discrepancy

An ongoing problem with Cepheid modeling is that masses of Cepheids determined using orbital dynamics of a stellar companion (see, e.g., Evans et al. 2018) or determined from pulsation modeling (see, e.g., Caputo et al. 2005) are smaller than those determined from standard stellar evolution models, with the discrepancy being as much as 20%, and being larger for Cepheids with smaller masses and shorter periods. Figure 5, adapted from Evans et al. (2018), shows evolution tracks for 4, 5 and 7 $M_{sun}$ models along with the positions of Polaris and V1334 Cyg (Gallenne et al. 2018) and their masses determined from binary orbital dynamics. The dashed lines correspond to evolution models calculated with rotation rates at 95% of breakup velocity.

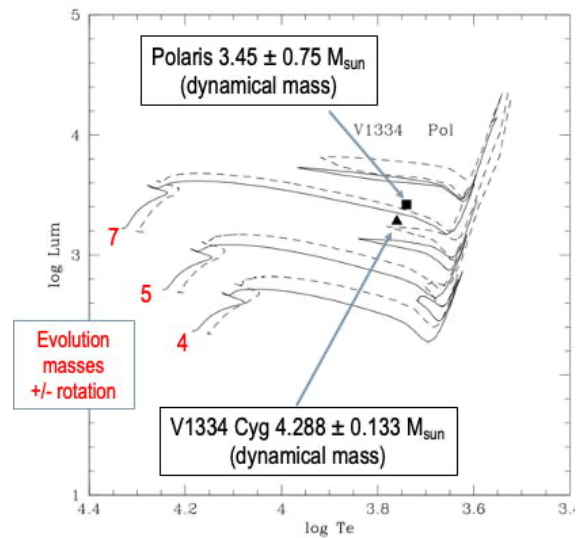

Figure 5. H-R diagram showing evolution tracks calculated without (solid) and with (dashed) rotation for 4, 5, and 7 $M_{sun}$ models, along with the locations of Cepheids Polaris and V1334 Cyg and their masses determined from binary orbital dynamics. Adapted from Evans et al. 2018; © AAS. Reproduced with permission.

One proposed method to mitigate the mass discrepancy is to include *core convective overshooting* in the models (see, e.g., Neilson et al. 2011). The cores of Cepheids were convective while they were burning hydrogen on the main sequence and Cepheids also have convective cores during the helium-burning phase. Overshooting at convection-zone boundaries can mix additional fuel into the core, extending the helium-burning phase and resulting in a larger core and higher luminosity. However, there are many proposed core overshooting prescriptions, and the amount of overshooting needed to solve the mass discrepancy problem is extreme if relying on this modification alone.

Another proposed method is to include *rotation* in the models (Anderson et al. 2014). Like convective overshooting, rotation can cause additional mixing between the core and envelope and increase the stellar luminosity. However, like core overshooting, high rates of rotation are needed to solve the problem, and models with too-high rotation rates result in higher-than-observed abundances of heavier elements mixed to the surface at the end of the main sequence or 'dredged-up' by envelope convection as the star becomes a red giant.

Many Cepheids are members of binary or multiple star systems (Neilson et al. 2015). Some Cepheids may have had a close companion at formation, and the two stars may have exchanged mass or merged together when one of the stars became a red giant and a common envelope formed. The resulting star may have an increased core size and luminosity-to-mass ratio compared to a single star that had evolved normally.

*Mass loss* via winds enhanced by pulsation has also been proposed to solve the problem (Neilson et al. 2011). However, the mass-loss rates required are as high as $10^{-7}$ $M_{sun}$/year, higher than observations for most Cepheids. For comparison, the Sun's mass-loss rate via the solar wind is 2 x $10^{-14}$ $M_{sun}$/year.

*Nuclear reaction rates* for helium burning (the triple-$\alpha$ process, where $\alpha$ represents the helium nucleus), as well as for $^{12}C(\alpha,\gamma)^{16}O$ reactions, which are both occurring in Cepheid cores. These reactions cannot be measured in the laboratory, are based on theory, and are believed to be uncertain by as much as 50% (Fields et al. 2018; Farag et al. 2022). However, an increase in these rates of more than a factor of 3 is needed to increase the luminosities of Cepheids enough to resolve the Cepheid mass problem.

Figure 6 shows the evolutionary tracks for models with and without a factor of 3 increase in these two reaction rates (Guzik et al. 2020). These evolution tracks were calculated using the MESA code (see section 3) and include a standard prescription for convective overshooting.

Another ingredient in stellar models, *opacities*, which regulate the transport of energy from the core, are also based on theory with some benchmarks from laboratory experiments. In the early 1990s, an increase in opacities mostly solved the 'beat and bump' Cepheid mass discrepancy, in which period ratios of double-mode Cepheids (first-overtone to fundamental or 2nd-overtone to fundamental) did not agree with those of pulsation models (Moskalik et al. 1992; Simon 1995).

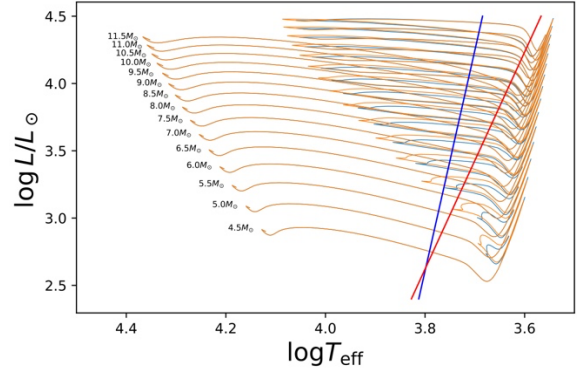

Figure 6. MESA evolution tracks for models without (blue) and with (yellow) a factor of 3 increase in the triple-$\alpha$ and $^{12}C(\alpha,\gamma)^{16}O$ nuclear reaction rates. The boundaries of the Cepheid instability region (red and blue lines) are from Bono et al. (2000).

Over the past two decades, further changes in opacities have been proposed to explain the 'solar modelling problem' (see, e.g., Buldgen et al. 2019), in which the internal sound speed in solar models differs from that inferred from solar oscillations. Opacity modifications are proposed by Daszyńska-Daszkiewicz et al. (2017) to explain the observed pulsation frequencies of ν Eridani, a main-sequence variable showing acoustic modes characteristic of beta Cep variables, and gravity modes characteristic of SPB (slowly pulsating B) stars. These types of stars are in the mass range that we expect will become Cepheids later in their lives.

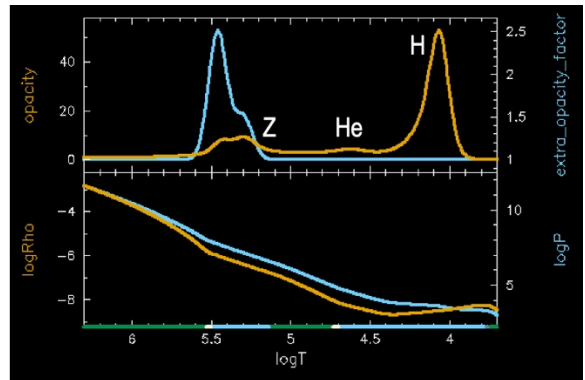

Figure 7. Opacity and opacity factor (top), density (Rho) and pressure (P) (bottom) vs. interior temperature for MESA calculation of an 8 $M_{sun}$ model within the Cepheid instability strip including enhanced opacities proposed to explain the pulsations of beta Cep and SPB stars.

We used the MESA code to calculate the evolution and pulsations of models including the proposed opacity enhancements (Guzik et al. 2020). The enhancement mainly affects the layers just below the so-called 'Z bump' around 200,000 K in the stellar envelope caused by atomic transitions of elements near Fe in the periodic table. This Z bump is responsible for the pulsations of the beta Cep and SPB stars.

Figure 7 shows the relative opacity bump sizes for an 8 $M_{sun}$ model in the Cepheid pulsation instability region and the magnitude and shape of the opacity multiplier function, which is as large as x 2.5. Figure 8 shows evolutionary tracks for 4.5 to 11.5 $M_{sun}$ models with and without the opacity enhancement, again using nominal convective overshooting settings. The opacity modification hardly affects the evolutionary tracks and only slightly decreases the extent of blue loop excursions.

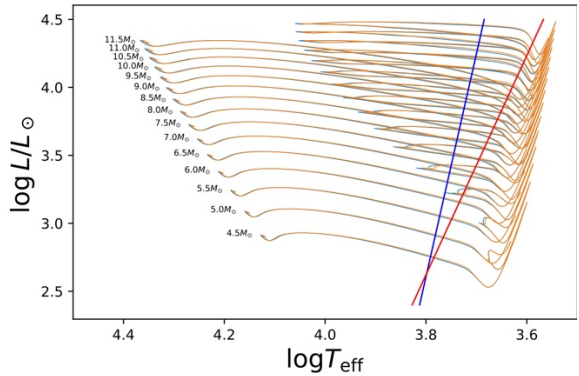

Figure 8. MESA evolutionary tracks for models without (blue) and with (orange) opacity enhancement proposed by Dazsyńska-Daszkiewicz et al. (2017) to explain the pulsation spectrum of ν Eri, a beta Cep/SPB main-sequence pulsator.

Dazsyńska-Daszkiewicz et al. (2023) found that period ratios of double-mode high-amplitude delta Scuti stars cannot be matched with stellar models for two out of three sets of commonly used stellar opacities. This problem should also affect RR Lyrae and Cepheid pulsators which are driven by the same 'kappa effect' opacity mechanism as the delta Scuti stars. However, at present there is no compelling atomic physics motivation for modifying the opacities.

It may be the case that some combination of stellar physics modifications will result in a physically reasonable solution to the Cepheid mass discrepancy. Finding the actual solution is likely to require careful work comparing various types of observations with evolution and pulsation models for different types of stars across the H-R diagram to improve the modeling of these processes and constrain their parameters and ranges. For example, Pedersen et al. (2021) used gravity-mode pulsations of SPB stars to constrain internal mixing processes.

## 3. MESA Models

During the past several years we have made use of the open-source Modules for Experiments in Stellar Astrophysics (MESA) code (Paxton et al. 2019; Jermyn et al. 2021) to explore Cepheid evolution and pulsation modeling. MESA can be installed on desktop and laptop computers, is well supported, and contains many tutorials and examples. MESA can be used to generate evolution models through the Cepheid phase, and to experiment with varying physics inputs such as convection treatment, opacities, and reaction rates, as illustrated in section 2. In 2019, a separate radial stellar pulsation (RSP) module was added to MESA which allows one to calculate radial pulsation hydrodynamics for an envelope-only model using a time-dependent convection treatment.

In addition to the Cepheid studies presented in Paxton et al. (2019), examples of other studies making use of MESA for Cepheids can be found in Das et al. (2020) and Gautschy (2019).

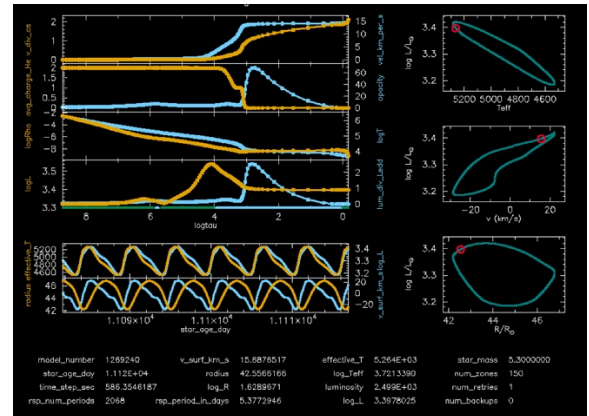

Figure 9. MESA RSP 'dashboard' showing various properties of δ Cep model during pulsation cycles.

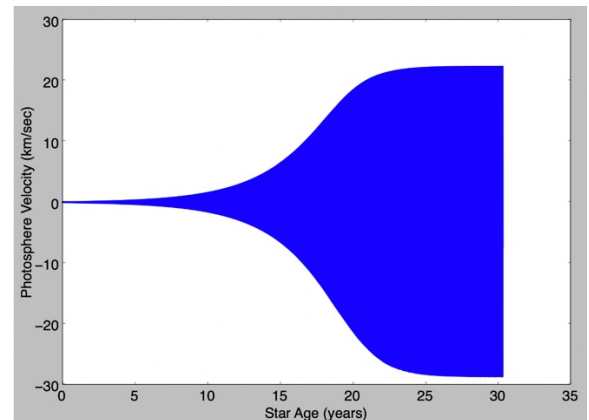

Figure 10. MESA RSP radial velocity vs. time for δ Cep model.

We applied MESA RSP to models for δ Cep, V1334 Cyg, and Polaris (Guzik et al. 2020). δ Cep pulsates in the radial fundamental mode with period 5.36 days, varies between 3.48 and 4.37 V magnitude, and shows radial velocity variations with amplitude around 20 km/sec (Stevenson 2006). We chose a mass of 5.3 $M_{sun}$ for our δ Cep model, and we estimated a luminosity (L) and effective temperature ($T_{eff}$) from literature values. We then varied $T_{eff}$, but also L, if necessary, until we found a model with predicted fundamental mode period 5.36 days. The resulting model has L=2089 $L_{sun}$ and $T_{eff}$ = 5861 K. We then initiated the model in the fundamental mode at 0.1 km/sec. Figure 9 shows a MESA 'dashboard' of model properties that can be monitored during the hydrodynamic pulsation calculations. After around 25 years of star time, the model converges to a radial velocity amplitude +20/-30 km/sec, in good agreement with observations (see Fig. 10).

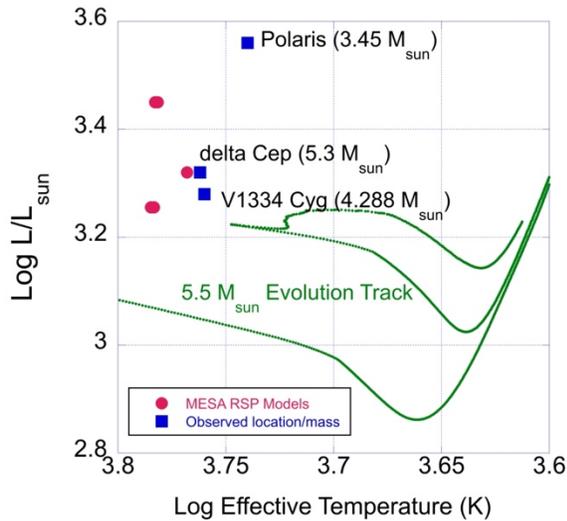

Figure 11. H-R diagram showing a 5.5 $M_{sun}$ MESA evolution track, observed locations and masses of Polaris, δ Cep, and V1334 Cyg, and locations of MESA RSP pulsation models (Guzik et al. 2020).

Figure 11 shows the H-R diagram for a 5.5 $M_{sun}$ MESA evolution track, along with the observed locations of Polaris, δ Cep, and V1334 Cyg, and the best-fit MESA RSP pulsation models (Guzik et al. 2020). For Polaris, we could not find a pulsating model using the dynamical mass 3.45 $M_{sun}$ and instead needed to increase the mass to 5.93 $M_{sun}$ to match the observed pulsation period. These envelope-only pulsation models all lie well above the 5.5 $M_{sun}$ evolution track, confirming the Cepheid mass discrepancy.

## 4. 2-D and 3-D Models

Most Cepheid pulsation and evolution models are one-dimensional (1-D), i.e., the model is assumed spherically symmetric and zones are spherical shells along the radial direction only. Advances in computational capabilities are allowing 2-D and 3-D models to be explored. Some of these models calculate a solid angle slice of the model and include only the outer portion in radius to save computational time mitigate time step problems associated with large disparities in timescale between various physical processes in the core vs. the envelope vs. the atmosphere.

Geroux and Deupree (2011, 2013, 2014, 2015): calculated RR Lyrae pulsations with Large Eddy Simulation (LES) convection, finding subtle differences in the averaged light curves between 2-D and 3-D simulations, with the 3-D light curves showing a more pronounced 'bump' feature.

Mundprecht et al. (2013, 2015) developed and applied the ANTARES code to evaluate non-local time-dependent convection treatments for 2-D Cepheid models and attempt to develop a realistic 1-D approximation that could be implemented in codes such as MESA.

Vasilyev et al. (2017, 2018, 2019) use CO5BOLD to evaluate spectral properties of 2-D Cepheid models, in particular to highlight radiative and convective flux transport variations in space and time for pulsating models.

Pratt et al. (2020) have been applying the 2-D/3-D Multidimensional Stellar Implicit Code (MUSIC) to full-sphere simulations of pre-main sequence stars and red giants incorporating 90% of the stellar radius. Animated visualizations (Pratt 2021) can be found at: https://doi.org/10.6084/m9.figshare.14128619.v1  We hope to collaborate with Pratt to apply this code to Cepheid models.

## 5. Multiwavelength Observations

Cepheid observations in ultraviolet (UV), X-ray, and radio wavelengths are providing new challenges for modeling. Engle et al. (2017) compare observations of δ Cep in the optical, far ultraviolet, and X-ray wavelengths, finding a spike or flare feature in the X-ray light curve at phase near 0.5, suggesting a new class of X-ray variable star (Figure 12). Multi-dimensional modeling has also been done to explore possible origins of such features (Moschou et al. 2020).

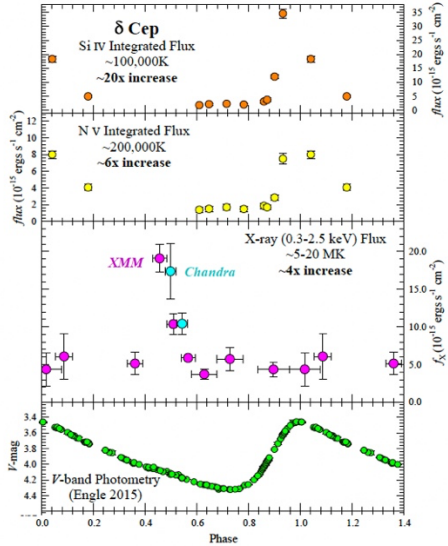

Figure 12. Observations of δ Cep in far-UV (top), X-ray (middle), and V-band (bottom) vs. light curve phase. Adapted from Engle et al. (2017). © AAS. Reproduced with permission.

Attempts to find X-ray bursts in other Cepheids and correlate with phase are ongoing (Evans et al. 2022). X-ray observations of η Aql, with very similar period and therefore luminosity and mass to δ Cep, do not show an X-ray burst, so a simple relationship between X-ray variability and pulsation phase does not seem universal.

δ Cep has also been detected in radio wavelengths at the VLA (Matthews et al. 2023). Spectro-polarimetric measurements detecting magnetic signatures across all pulsation phases of bright Galactic Cepheids Polaris, ζ Gem, η Aql, and δ Cep (Barron et al. 2022) show promise for advancing understanding of Cepheid atmospheres.

## 6. TESS Observations

The NASA Transiting Exoplanet Search Satellite (TESS; Ricker et al. 2015) was launched in 2018 to discover new planets in the habitable zone around nearby stars. The spacecraft is also acquiring high-precision, long-time series, high-cadence photometry for many other objects in the observing fields. The observations are taken in fields called 'sectors' for 27 days each. 2-minute cadence or even 20-second cadence data are available. We applied to the TESS Guest Observer program in Cycle 3 (targeting fields south of the ecliptic plane) and Cycle 4 (observing north of the ecliptic plane) to monitor bright Galactic Cepheids chosen from the David Dunlop Observatory catalog (see section 1). One of our goals is to determine whether we could detect features in the TESS passband (600-1000 nm centered on Cousins I) at certain phases that might correspond to UV or X-ray features as seen in δ Cep.

We now have around sixty 2-minute cadence light curves from Cycle 3 and forty 2-minute cadence light curves for Cycle 4 that we are in the process of analyzing. We have chosen several from Cycle 3 to show (Figures 13-17). The light curves show interesting diversity, which will provide many challenges to modelers.

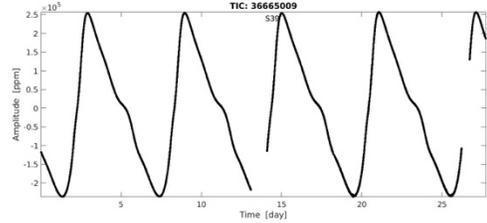

Figure 13. TESS light curve of RV Sco (TESS mag 6.01) observed in sector 39 (May 27-June 24, 2021).

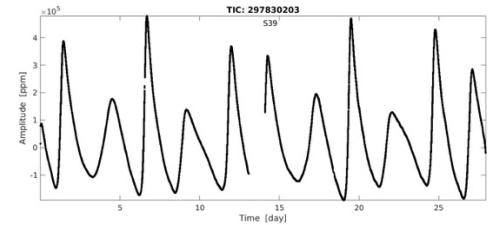

Figure 14. TESS light curve of U TrA (TESS mag 8.02) observed in sector 39 (May 27-June 24, 2021).

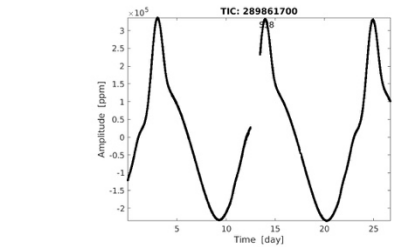

Figure 15. TESS light curve of XX Cen (TESS mag 6.97) observed in sector 38 (April 29-May 26, 2021).

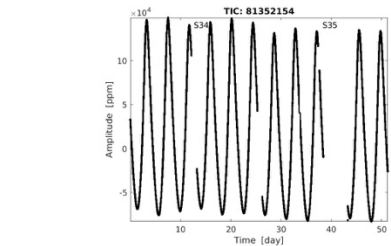

Figure 16. TESS light curve of AH Vel (TESS mag 5.17) observed in sectors 34 and 35 (January 14-March 6, 2021).

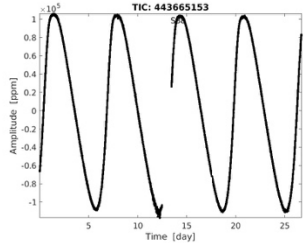

Figure 17. TESS light curve of V378 Cen (TESS mag 5.92) observed in sector 38 (April 29-May 26, 2021).

## 7. More Modeling Challenges

As in part illustrated by the TESS light curves, there remain many ongoing challenges for Cepheid modeling. Some of these include:
- Predicting Instability Strip Boundaries (see, e.g., Houdek and Dupret 2015).
- Explaining non-pulsating stars in the Cepheid instability strip (20-30% of stars in the instability strip in the LMC do not show pulsations; Narloch et al. 2019).
- Creating double-mode Cepheids in nonlinear hydrodynamic models--the models tend to settle on or switch to a preferred mode (see, e.g., Fadeyev 2021; Buchler 2009).
- Understanding period changes and correlating with evolution phase and crossings of H-R diagram. Period changes are fast on the first crossing of instability strip toward the RGB, change sign and are slower on the second crossing, and change sign again on the third crossing as the star moves toward the AGB (see, e.g., Espinoza-Arancibia, et al. 2022; Miller et al. 2020; Fadeyev 2014).
- Understanding and modeling non-radial pulsation modes observed in some Cepheids (see, e.g., Smolec 2019; Buchler 2009).
- Explanation of amplitude changes and 'Blazhko effect' as also seen in RR Lyrae stars (see, e.g., Szabó 2013; Plachy et al. 2021).

## 8. Acknowledgements

J.G. thanks the American Physical Society for an invitation to present a review talk at their April 2023 meeting, providing the impetus for this paper. J.G. also thanks collaborators from the MESA Summer School 2019: Ebraheem Farag, Jakub Ostrowski, Bill Paxton and Frank Timmes, Cepheid collaborators Hilding Neilson, Sofia Moschou, Jeremy Drake, Scott Engle, Stephanie Flynn, Jane Pratt, and Karen Kinemuchi, and LANL collaborators Philipp Edelmann and Melissa Rasmussen. We are grateful to the TESS Guest Investigator Program for high-quality Cepheid data. J.G. is grateful for support from Los Alamos National Laboratory, operated by Triad National Security, LLC, for the Department of Energy's National Nuclear Security Administration, Contract #89233218CNA000001.